\documentclass[12pt,english,british]{iopart}
\usepackage[T1]{fontenc}
\usepackage[latin9]{inputenc}
\usepackage{babel}
\usepackage{textcomp}
\usepackage{amssymb}
\usepackage{graphicx}
\usepackage[unicode=true,
 bookmarks=true,bookmarksnumbered=false,bookmarksopen=false,
 breaklinks=false,pdfborder={0 0 1},backref=false,colorlinks=false]
 {hyperref}
\hypersetup{pdftitle={Positive streamer inception and propagation in high-purity nitrogen: effects of voltage rise-rate},
 pdfauthor={S. Nijdam}}
\usepackage{breakurl}

\makeatletter

\providecommand{\tabularnewline}{\\}

\usepackage{iopams}
\usepackage{setstack}

\usepackage{babel}

\usepackage{tabu, caption}

\makeatother

\begin{document}

\title[Investigation of streamers by double pulse experiments]{Investigation of positive streamers by double pulse experiments,
effects of repetition rate and gas mixture}

\author{S.~Nijdam$^{1}$, E.~Takahashi$^{2}$, A.H. Markosyan$^{3}$ and
U.~Ebert$^{1,3}$}

\address{$^{1}$ Eindhoven University of Technology, Dept.\ Applied Physics\\
 P.O. Box 513, 5600 MB Eindhoven, The Netherlands\\
$^{2}$ National Institute of Advanced Industrial Science and Technology
(AIST),\\
1-2-1 Namiki, Tsukuba Ibaraki 305-8564, Japan}

\address{$^{3}$ Centrum Wiskunde \& Informatica (CWI), Amsterdam, The Netherlands}

\ead{s.nijdam@tue.nl}
\begin{abstract}
Streamer discharges are often operated in a repetitively pulsed mode
and are therefore influenced by species left over from the previous
discharge, especially free electrons and ions. We have investigated
these effects by applying two consecutive positive high voltage pulses
of 200--700\,ns duration to a point-plane gap in artificial air,
pure nitrogen, other nitrogen-oxygen mixtures and pure argon at pressures
between 67 and 533\,mbar. The pulses had pulse-to-pulse intervals
($\Delta t$) between 200\,ns and 40\,ms. We imaged both discharges
with two ICCD cameras and combined this to a compound image. We observe
for values of $\Delta t$ below 0.5--15\,\textmu{}s (at 133\,mbar,
depending on gas mixture) that during the second pulse the streamers
continue the paths of the first-pulse streamers. We call the maximal
time for which this continuation still occurs the continuation time.
For N$_{2}$-O$_{2}$ mixtures, this time has a maximum at an oxygen
concentration of about 0.2\,\%. According to our plasma chemical
modelling this maximum is determined by the electron loss rate which
has a minimum around this oxygen concentration. Depending on oxygen
concentration the dominant recombining positive ion is N$_{4}^{+}$,
O$_{2}^{+}$ or O$_{4}^{+}$ where O$_{2}^{+}$ dominates around 0.2\,\%~O$_{2}$
and recombines slowest.

For increasing values of $\Delta t$ we observe that after the continuation
phase first no new streamers occur at all, then new streamers show
up that avoid the entire pre-ionized region. Next we see new thin
streamers that follow the edges of the old channels. For larger $\Delta t$
(10--200\,\textmu{}s) the new streamers start to increase in size
and move to the centre of the old channels. Finally, around millisecond
timescales the new channels are completely independent of the old
channels. 

Together this points to the combination of two mechanisms: streamers
search the proximity of regions with increased electron density, but
cannot penetrate regions with too high electron density.
\end{abstract}

\submitto{\PSST}

\maketitle

\section{Introduction\label{sec:Introduction}}

It has been observed by multiple groups that streamer discharges are
in many ways influenced by preceding discharges~\cite{Nijdam2011c,Nijdam2011,Shao2006,Pai2010}.
From these and other studies, it is clear that species produced by
such discharges play an important role in inception and propagation
of streamers. In most cases it is assumed that the dominant effect
of preceding discharges is the creation of (background) ionization.
Immediately after a discharge the background ionization consists primarily
of free electrons and positive ions, depending on composition and
pressure. For longer timescales this generally changes to positive
and negative ions due to electron attachment and/or recombination,
again depending on composition and pressure. When the positive and
negative species are not distributed equally this will lead to space-charge
which can also influence subsequent streamer discharges by its resulting
electric field. Other species that can play a role besides charged
species are radicals (e.g. N and O), metastables (e.g. N$_{2}\mathrm{(A^{3}\Sigma_{u}^{+})}$
and Ar(4s)) as well as other products like NO and OH.

One way of studying the effects of initial conditions on a streamer
discharge is to create background ionization with an independent source
like a laser~\cite{Takahashi2011}, x-rays~\cite{Mathew2007} or
radioactive compounds~\cite{Nijdam2011c}. This has the advantage
that it decouples the production of ionization (and other important
long-lived species) from the streamers. 

Here, we will investigate the effects of a preceding streamer discharge
on a subsequent discharge by applying two positive high voltage pulses
in succession with pulse-to-pulse intervals ($\Delta t$) between
200\,ns and 40\,ms. This allows us to study the effect of the preceding
discharge on a very large range of time scales. At low values of $\Delta t$
the time scale is of similar order as the characteristic time of the
total streamer development and the duration of the applied pulse.
Therefore we can also learn things about streamer development in general.
Values of $\Delta t$ around 100\,\textmu{}s to 1\,ms correspond
to repetitive streamer discharges with 1 to 10\,kHz repetition frequency
that are often used in commercial streamer and plasma-jet applications~\cite{Winands2006a,Laroussi2007,Starikovskiy2013}.
Finally, at $\Delta t\geq10$\,ms the discharges start to resemble
the 'single pulse' discharges often used to study streamer development.

The advantage of applying two pulses instead of a continuous pulse
train is that it is much easier to see the effects of a single pulse
discharge on the next discharge and that it allows the huge range
of $\Delta t$ that was discussed above. It is experimentally extremely
challenging to build a pulse source with a variable repetition frequency
between 10\,Hz and 50\,MHz while maintaining the same pulse shape
and amplitude. Creation of two identical pulses with variable interval
is much easier.\\

Walsh \emph{et al}~\cite{Walsh2012,Naidis2013} have shown in experiments
with atmospheric pressure plasma-jets in helium that, for jets extending
beyond the guiding tube, the propagation can be temporarily halted
by 'neutralizing' the applied voltage by applying the same voltage
to two external electrodes surrounding the jet. This resembles the
application of two consecutive pulses to a single electrode like in
our experiments. \\

In this paper we will discuss the following topics: in section~\ref{sec:Experimental-setup}
we give the details and parameters of our experiment and methods.
Section~\ref{sec:General-development} treats the general development
of streamer discharges created with two pulses as function of the
time gap $\Delta t$ between pulses by means of ICCD imaging. This
is done for three gases: artificial air, pure nitrogen and pure argon
for various pressures and pulse widths and amplitudes. In section~\ref{sec:Duration-of-streamer}
we focus on the value of $\Delta t$ for which we can still observe
continuation of the first-pulse streamers during the second pulse
and how this depends on gas conditions, most notably on the variation
of the oxygen concentration in nitrogen-oxygen mixtures. The experimental
results are then compared to plasma-chemical modelling results in
section~\ref{sec:Modelling}. Finally we will summarize our results
and draw conclusions in section~\ref{sec:Discussion-and-conclusions}.

\section{Experimental setup\label{sec:Experimental-setup}}

\begin{figure}
\centering\includegraphics[width=8cm]{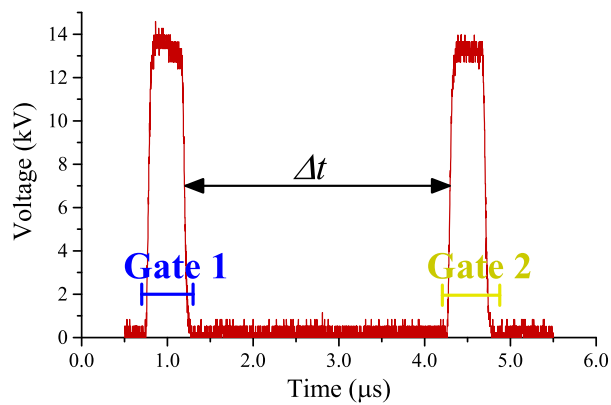}

\caption{\label{fig:Voltage-shape}Typical shape of the two subsequent voltage
pulses with the pulse-to-pulse interval \foreignlanguage{english}{$\Delta t$}
indicated. Indications of the opening (gate) times of the two different
ICCD cameras are shown in blue and yellow respectively.}
\end{figure}

All experiments are performed in a 103\,mm point-plane gap mounted
inside a vacuum vessel. This vessel is filled with artificial air,
pure nitrogen, pure argon and air-nitrogen mixtures at pressures between
67 and 533\,mbar. When using pure nitrogen or argon, the vessel is
flushed at a rate of roughly 1 standard litre per minute to improve
gas purity in the experiments. The purity of all of the source gases
is specified as 6.0 (less than 1\,ppm impurities). However, because
of leakage and/or out-gassing, we estimate the impurity level during
experiments at about 5\,ppm%
\footnote{This was calculated by dividing the measured leak/out-gassing rate
of $5\cdot10^{-6}$~standard litres per minute by the applied gas
flow of 1 standard litre per minutes.%
}. On the pointed tungsten tip with a tip radius of 15\,\textmu{}m,
a pair of positive voltage pulses is applied repeatedly. Such a pulse
pair consists of two nearly identical pulses with pulse-to-pulse intervals
($\Delta t$) between 200\,ns and 40\,ms. The pulse pairs are applied
with a repetition frequency of 0.7\,Hz. The individual pulses that
make up the pulse pairs are relatively rectangular shaped with pulse
lengths of 200--700\,ns, rise/fall times of about 15\,ns and pulse
amplitudes of 7--17\,kV. An example of such a pulse-pair is shown
in figure~\ref{fig:Voltage-shape}, where the definition of $\Delta t$
is indicated as well. Pulse amplitudes are here defined as the average
voltage of the plateau during the first pulse. For small values of
$\Delta t$ the amplitude during the second pulse can be slightly
smaller as is shown in figure~\ref{fig:Voltage-shape}.

\begin{figure}
\centering\includegraphics[width=8cm]{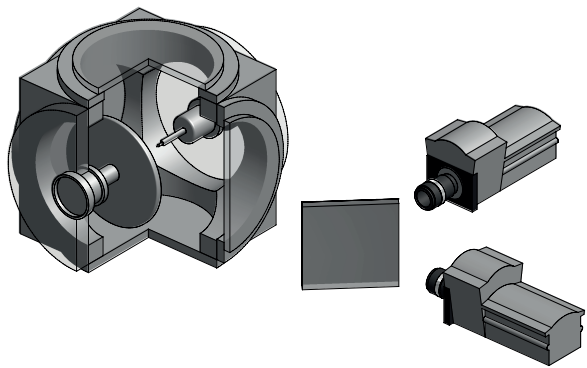}

\caption{\label{fig:Setup}Schematic illustration of vacuum vessel, half-mirror
and two ICCD-cameras. Part of the vacuum vessel has been cut from
the rendering to show the tip and plate. In all presented images the
recorded images are rotated anti-clockwise by 90$^{\circ}$ so that
the electrode tip is shown on top.}
\end{figure}

The streamer-discharges induced by the two voltage pulses are imaged
by two separate, nearly-identical ICCD cameras. These cameras are
mounted so that they image exactly the same plane, by means of a half-mirror
(see figure~\ref{fig:Setup}). During post-processing the images
are rotated so that the electrode tip is always located at the top
of the image. The gate of one camera is opened during the first voltage
pulse; the gate of the other camera is opened during the second voltage
pulse (see gate indications in figure~\ref{fig:Voltage-shape}).

\begin{figure}
\centering\includegraphics[width=8cm]{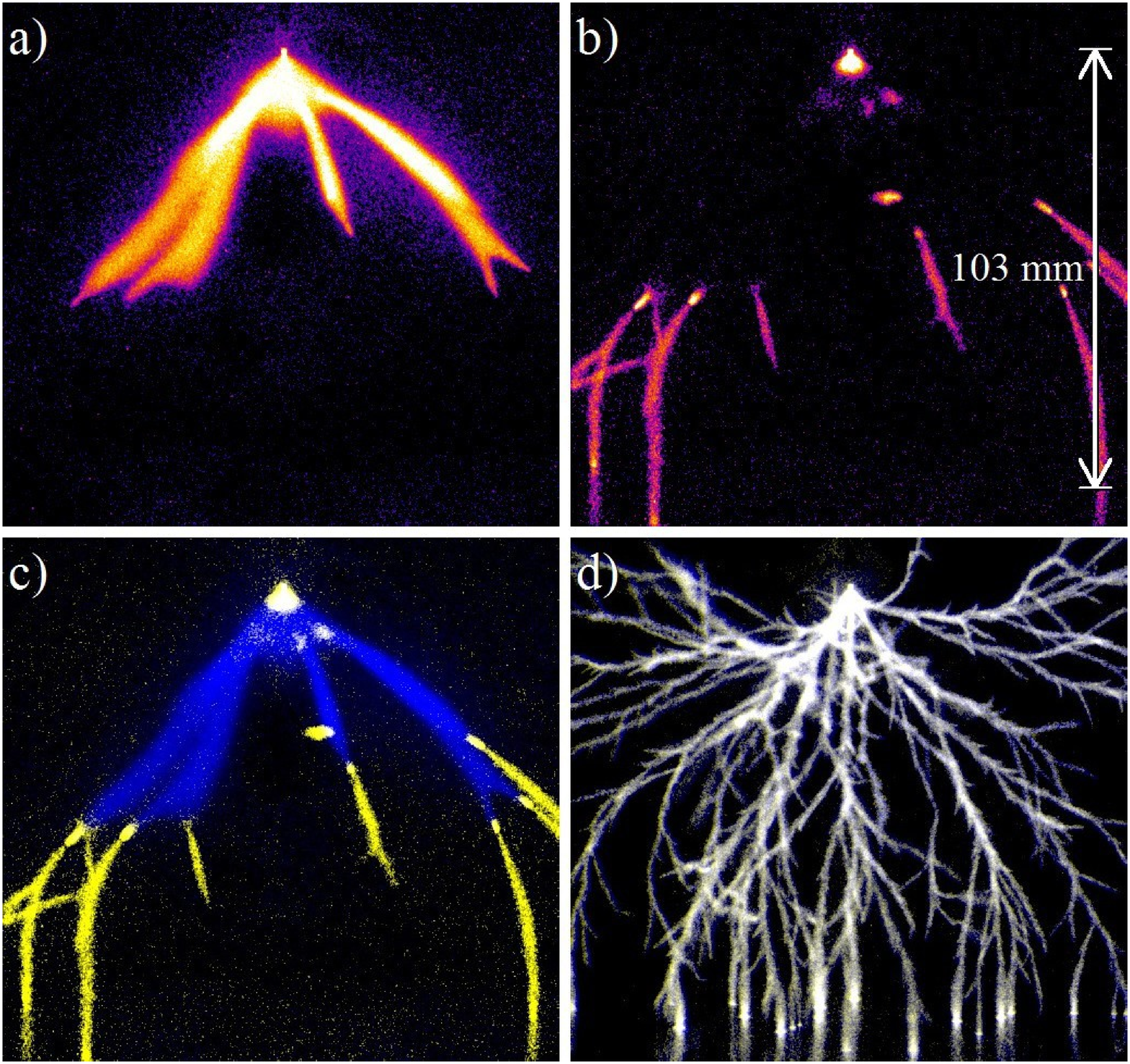}

\caption{\label{fig:Illustration-of-measurement}Illustration of the two-camera
overlay technique. a) and b) show false-colour representations of
the captured frames from camera 1 and 2 respectively, recorded during
one voltage pulse pair (conditions: 6.8\,\% O$_{2}$ in N$_{2}$,
133\,mbar gas, 13.6\,kV, 200\,ns pulses with $\Delta t=500$\,ns)
. In c) the images from a) and b) are superimposed in blue and yellow
respectively. d) shows a similar overlay of a discharge under different
conditions, with both camera gate times set at the first pulse only
(conditions: pure N$_{2}$, 133 mbar gas, 13.6\,kV, 400\,ns pulse).}
\end{figure}

A typical example of an image-pair produced by the two cameras is
shown in figures~\ref{fig:Illustration-of-measurement}a and~b,
where the images are acquired during the first and second pulse respectively.
These two images are then laid over each other as a new image where
the blue channel (component) of the RGB-pixels represents the first
pulse image and the red and green channels represent the second pulse
image. A pixel with the red and green channels equally bright and
the blue channel off will appear yellow, while a pixel with all three
channels fully illuminated will appear white. Therefore areas that
only emit during the first pulse are blue, areas that only emit during
the second pulse are yellow and areas that emit during both pulses
are white. All linear combinations of yellow and blue are possible.
The result of this operation on figures~\ref{fig:Illustration-of-measurement}a
and~b is shown in figure~\ref{fig:Illustration-of-measurement}c.

In the example from figures~\ref{fig:Illustration-of-measurement}a-c
it is clear that the (blue) streamer channels created by the first
pulse continue their path (in yellow) during the second pulse. The
area around the electrode tip, as well as some areas on the edge of
the inception cloud emit again during the second pulse and are therefore
coloured white. 

Figure~\ref{fig:Illustration-of-measurement}d shows a superimposed
image with both cameras gated around the first pulse (under different
conditions than in figures~\ref{fig:Illustration-of-measurement}a-c).
In this case both cameras should capture exactly the same image. This
shows that the two cameras are aligned properly as almost all streamer
channels are rendered as white or grey and only a slight yellow/blue
offset is visible.\\

Note that in all images presented in this paper the intensities of
both cameras are chosen arbitrarily. The same intensity on the image
does not imply that the original streamers have the same intensity.
This was done for several reasons: Firstly the two cameras are slightly
different and do not have the same gain factors. Secondly there was
a half-mirror between the vessel and the cameras which has different
losses for transmission and reflection and finally because the streamer
morphology is best imaged with the maximum contrast per image. For
these reasons we have used automatic contrast and brightness algorithms
to process the final output images. This may also explain the slight
blue tint in figure~\ref{fig:Illustration-of-measurement}d.

In all cases, observations are performed on two-dimensional images
of the three-dimensional discharge structures. Therefore it is not
always possible to draw conclusions on the exact relation between
first and second discharge from one image. To do this properly one
would need a stereoscopic set-up~\cite{Nijdam2008} to observe the
full three-dimensional relation between the paired images. Nevertheless,
observation of multiple images under the same conditions can give
good insight in the morphology and the relation between the paired
discharges. All observations discussed in this paper are based upon
multiple images per parameter setting (usually 5 or 10 images), but
only a limited subset of these images is presented here.

\section{\label{sec:General-development}General development}

We have studied the morphology of the discharge induced by the voltage
pulse pairs as function of pulse-to-pulse interval ($\Delta t$) for
a variety of gases, pressures and voltage pulse lengths and amplitudes.
In general the development during the second pulse as function of
$\Delta t$ is similar for all conditions, although there are variations
in details. We will first discuss the full development for one set
of conditions in air. Next we will look into the effects of pulse
length, voltage and pressure on this development. Finally we will
look at pure nitrogen and pure argon. 

Note that because the gate of the cameras is opened during the full
duration of the two respective pulses, propagating streamer heads
will be imaged as continuous trails~\cite{Ebert2006a,Marode2009,Nijdam2010}.
The trails created during the first pulse will indicate the position
of the channels of leftover species created during this pulse. For
sub-millisecond timescales the effects of diffusion and convection
on the location of the trails is expected to be minimal.

\begin{figure}
\includegraphics[width=16cm]{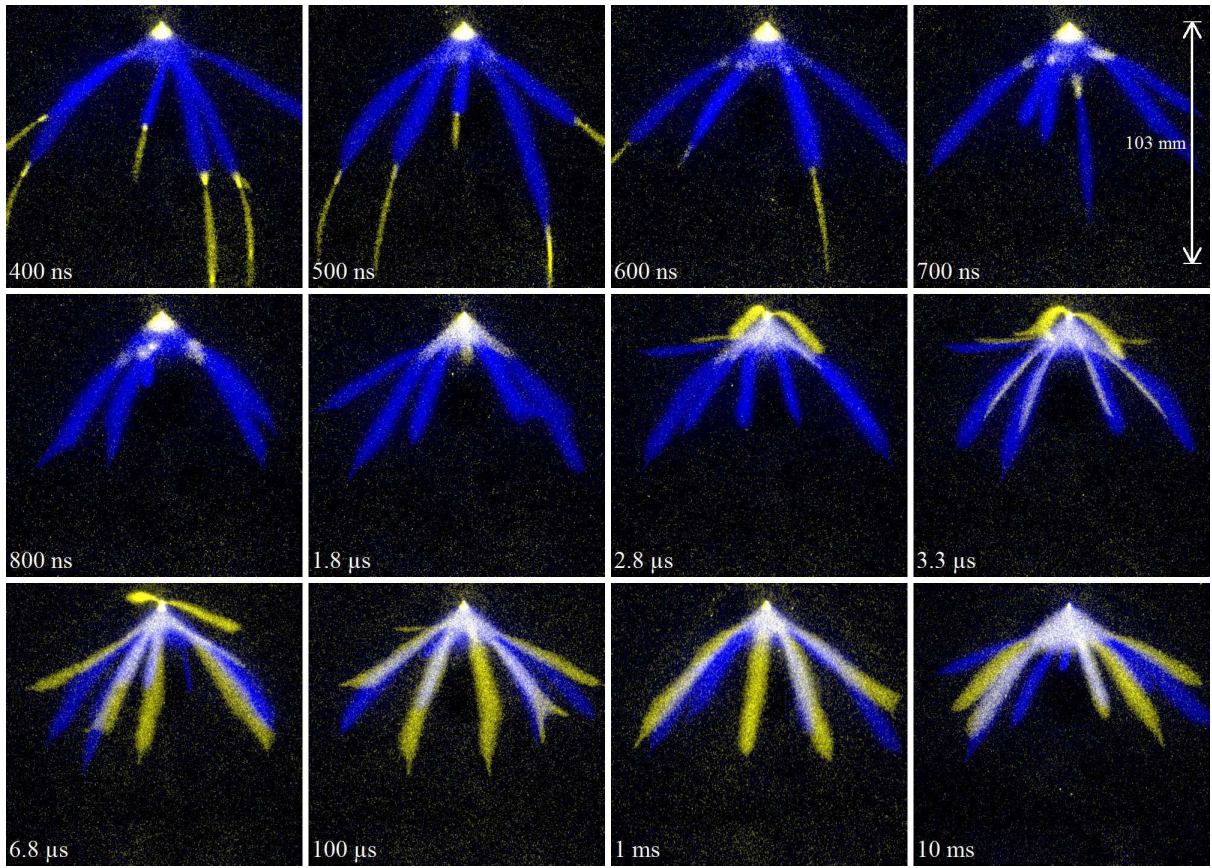}

\caption{\label{fig:Air-timeline}Superimposed discharge-pair images for varying
$\Delta t$ as indicated in the images. Images taken in 133\,mbar
artificial air with pulses of 13.6\,kV amplitude and 200\,ns pulse
length.}
\end{figure}

\subsection{Air\label{sub:Air}}

An example of the development of the discharges during the two pulses
in 133\,mbar air with relatively short pulses and a large variation
in $\Delta t$ is given in figure~\ref{fig:Air-timeline}. The discharges
generated by the first pulse are, except for stochastic variations,
similar for all values of $\Delta t$. These first-pulse discharges
consist of a so-called inception cloud that breaks up into a few (typically
five or six) streamer channels which mostly terminate either at the
moment the first pulse is stopped or when the (bottom) cathode plate
has been reached. In the conditions from figure~\ref{fig:Air-timeline}
the pulse duration has been chosen so that the streamers terminate
roughly halfway between electrode tip and cathode plate.

We can describe the development of the discharge during the second
pulse as function of $\Delta t$ for the conditions of ~\ref{fig:Air-timeline}with
the following six stages:
\begin{description}
\item [{(i)}] At very short pulse-to-pulse intervals ($\Delta t<700$\,ns),
the second-pulse streamers continue where the first-pulse streamers
stopped at the end of the first pulse. At the new starting point a
structure resembling a small scale inception cloud is visible. The
extended streamer channels become shorter, thinner and fewer for increasing
$\Delta t$. \\
Furthermore, the regions around the tip and the positions where the
first streamers originally emerged from the inception cloud both emit
light again, as can be seen from their white colour. 
\item [{(ii)}] At $\Delta t\ge700$\,ns the first-pulse channels are no
longer continued during the second pulse. Instead the re-glowing points
at the edge of the inception cloud become more pronounced and eventually
at $\Delta t$ = 1.8 \textmu{}s grow together with the emitting region
around the tip. 
\item [{(iii)}] At $\Delta t\ge2.5$\,\textmu{}s the first new streamer
channels start to appear. These channels avoid the entire old inception
cloud region and move around it. 
\item [{(iv)}] At $\Delta t=3.3$\,\textmu{}s more new channels appear.
These new channels seem to follow the edges of the old channels. They
are clearly thinner than both the old channels and the other new channels
that avoid the inception cloud. Note that at $\Delta t=1.8$\,\textmu{}s
a few short channels with similar morphology are also visible. Analysis
of a large number of images at these conditions has shown a preference
for the bottom side (with respect to the orientation of the presented
images) of the old streamer channels. Images made with shorter gate
times have revealed that these channels are the time-integrated representation
of travelling dots (streamer heads) and not continuous re-illumination
of existing channels.
\item [{(v)}] At larger values of $\Delta t$ the new channels increase
in width until they are roughly as wide as the original channels around
$\Delta t=5.8$\,\textmu{}s and start to overlap more with the parent
channels. For increasing $\Delta t$ they also start to become less
dependent on the location of the first-pulse streamers. However, up
to $\Delta t=1.0$\,ms a large number of the second-pulse streamers
still follows the first-pulse streamer paths, although not always
exactly. Under these conditions the inception cloud during the second
pulse is significantly smaller than during the first pulse. 
\item [{(vi)}] Around $\Delta t=10$\,ms the second discharge has become
fully independent of the first discharge. The inception cloud sizes
as well as the thickness and position distributions of the streamers
are now the same for the two pulses. Furthermore, the position of
the streamers during the second pulse no longer shows a correlation
with the positions during the first pulse. 
\end{description}
Note that these six stages cannot always be distinguished exactly
and that the exact transition between stages is often difficult to
pinpoint.

\paragraph{Interpretation}

The structures that are visible in stage (i) resemble the findings
of Walsh \emph{et al.}~\cite{Walsh2012} where atmospheric pressure
plasma-jets in helium also continue their path after the applied external
field was suppressed for a short period. We will discuss the stage
(i) streamer continuation and its interpretation in depth in sections~\ref{sec:Duration-of-streamer}
and~\ref{sub:Modelling-continuation}.

The re-emission of light from some regions during stages (i) to (iv)
is probably similar to a secondary streamer or glow discharge. This
is in fact not a real propagating streamer front but instead re-illumination
of a previously ionized area due to a change in field distribution.
A more detailed investigation has revealed that during the second
pulse the re-glowing of the points around the tip and the original
inception cloud occurs early in the pulse, while the continuation
of the old channels starts from about 50 to 100\,ns after the pulse
start. Note that it is very likely that parts of the old channels
that do not re-emit during the second pulse still carry (dark) current.
Because their conductivity is still high no high fields can exist
and current will not produce any light. For larger $\Delta t$ the
conductivity decreases and therefore the secondary streamers become
more pronounced.

The new channels avoiding the old ones that appear in stage (iii)
appear similar to the streamer morphology treated in~\cite{Nijdam2011},
where new streamers avoid the pre-ionized area generated by a previous
discharge. Similar behaviour was found by Briels \emph{et al.}~\cite{Briels2006}
where long voltage pulses were applied to a point plane gap in ambient
air and so-called late streamers are visible that avoid the area occupied
by earlier streamers. In both these cases as well as in the results
presented here the new streamers move around the area covered by the
preceding streamers or inception cloud. This is probably caused by
the high conductivity in this area which will quickly shield any electric
field and therefore inhibit the formation of an ionization front.
The position of the new channels at the edge of the old ones in stage
(iv) seems to be related to this: the inside of the old channels is
still too conductive for discharge propagation but the edges have
an electron density which is preferable for streamer propagation.
The preference for the bottom side of the channels may be due to the
higher background field in this region (closer proximity to the cathode
plane). Alternatively, as was proposed by Mark Kushner at the GEC
2013, the outer regions of the old channels may contain higher concentrations
of O$_{2}^{-}$ and may therefore be more attractive for streamer
propagation. In stage (v) the new channels can slowly penetrate the
area of the old channels due to the decrease in conductivity. The
guiding effect of the old channels on the new channels can also be
partly due to an increase in temperature, and therefore a decrease
in gas density, in the old channels. However, we do not expect a large
temperature increase~\cite{Riousset2010} so this effect will probably
be minor.

\begin{figure}
\centering\includegraphics[width=8cm]{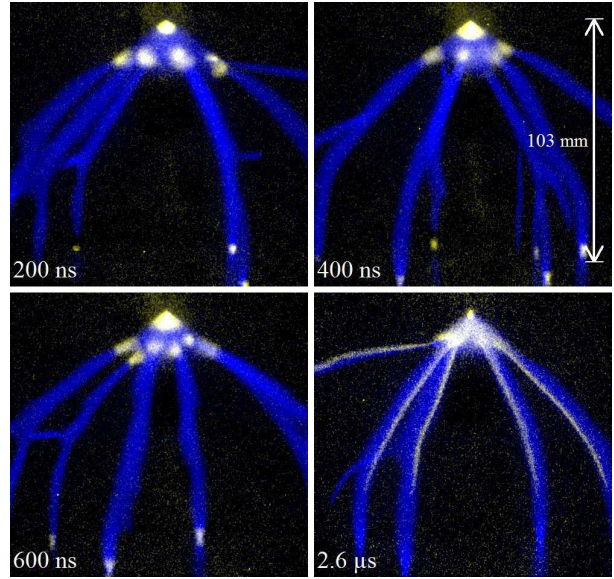}

\caption{\label{fig:AirLongerPulses}Superimposed discharge-pair images for
varying values of $\Delta t$. Images taken in 133\,mbar artificial
air with pulses of 13.6\,kV amplitude and 400\,ns length.}
\end{figure}

\paragraph{Longer pulses}

When 400\,ns voltage pulses instead of 200\,ns pulses are applied,
the development of the second-pulse streamer as function of $\Delta t$
is very similar (see figure~\ref{fig:AirLongerPulses}). The only
differences are related to the fact that the first-pulse streamers
have already crossed the gap. Therefore the occurrence of channel-continuation
for small values of $\Delta t$ (stage (i)) is rare as this only occurs
on the few branches that did not cross the channel during the first
pulse. However, the branches that did cross show something else: during
the second voltage pulse the cathode-spots (the \textquoteleft{}feet\textquoteright{}
of the channels) from the first discharge light up again. This effects
lasts to about $\Delta t=1$\,\textmu{}s. Furthermore, re-glowing
of the spots at the edge of the original inception cloud is more pronounced
with these longer pulses and the new channels start to occur for smaller
values of $\Delta t$ than with the shorter pulse (compare $\Delta t=2.6$\,\textmu{}s
from figure~\ref{fig:AirLongerPulses} to $\Delta t=2.8$\,\textmu{}s
from figure~\ref{fig:Air-timeline}).

\begin{figure}
\centering\includegraphics[width=16cm]{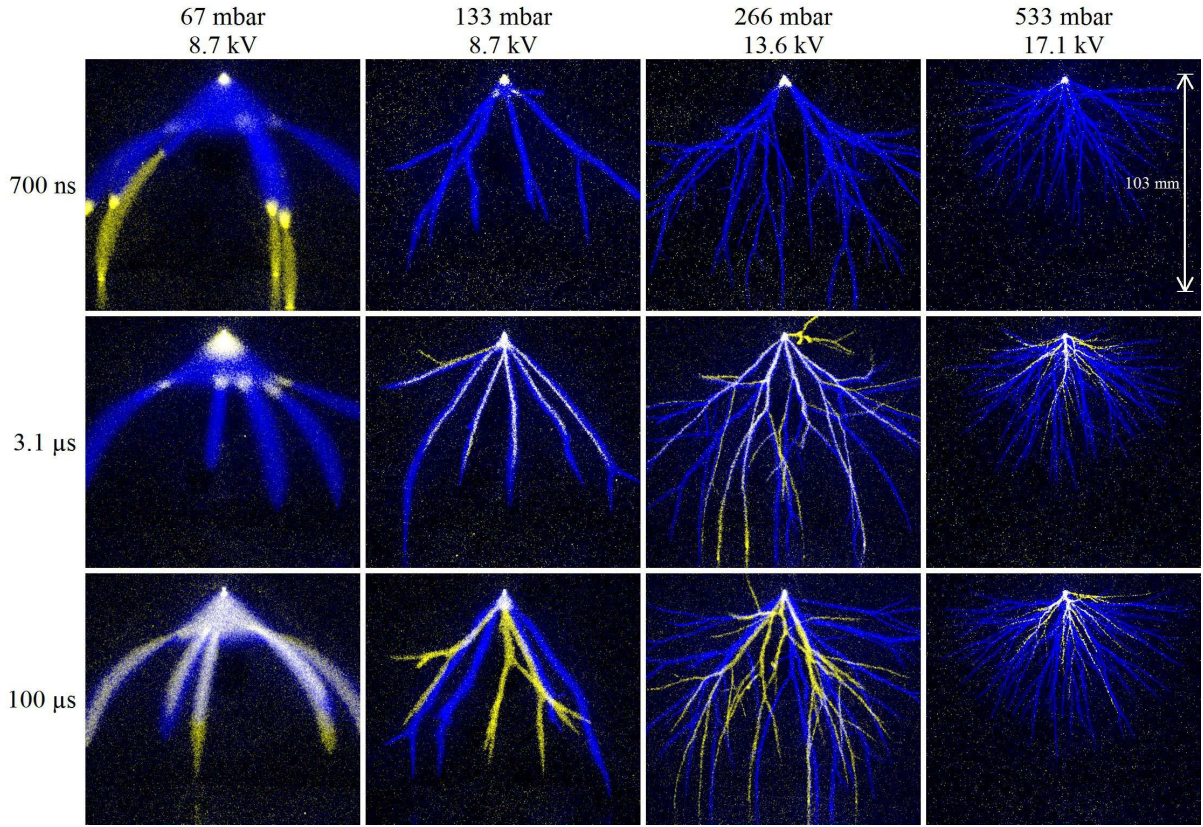}

\caption{\label{fig:Air-various-conditions}Superimposed discharge-pair images
for varying values of $\Delta t$ (indicated on the left), pressure
and voltage pulse amplitude (indicated on top). Images taken in artificial
air with pulses of 400\,ns length.}
\end{figure}

When changing voltage and pressure, the effects on the first and second
pulse morphologies are in general predictable (see figure~\ref{fig:Air-various-conditions}):
at higher voltages more and thicker streamers occur while at higher
pressures fewer and thinner streamers occur. Furthermore, the different
stages of development occur at lower values of $\Delta t$ for higher
pressures and vice-versa. More details on the effects of parameter
settings on streamer continuation will be given in section~\ref{sec:Duration-of-streamer}. 

\subsection{Pure nitrogen\label{sub:Pure-nitrogen}}

The same measurements as treated above on air have also been performed
on pure nitrogen. As was mentioned before, the expected purity in
pure nitrogen is about 5\,ppm. This purity is less than in previous
work~\cite{Nijdam2010} but still ensures an oxygen concentration
that is four to five orders of magnitude lower than in air.

\begin{figure}
\includegraphics[width=16cm]{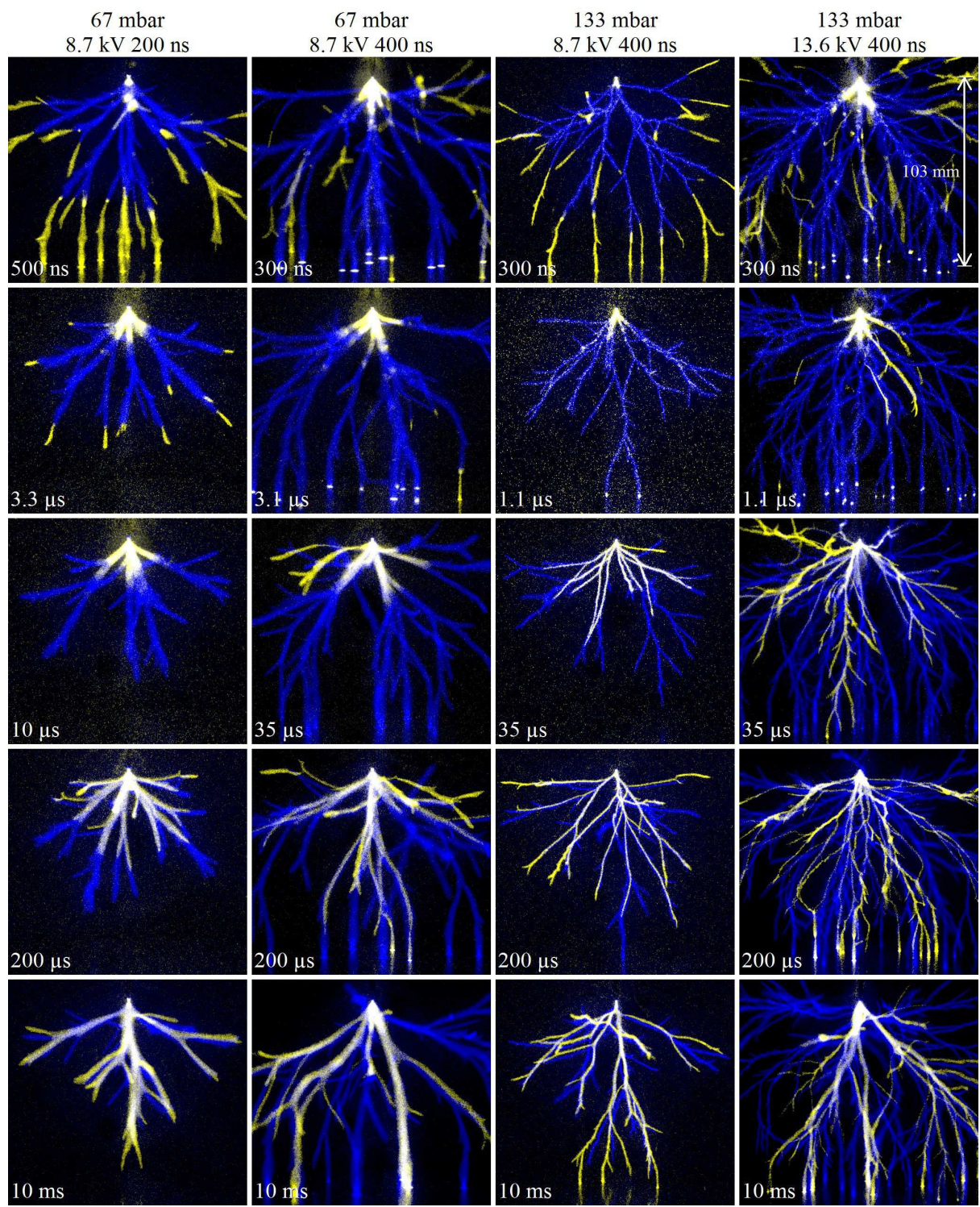}

\caption{\label{fig:Nitrogen-images}Superimposed discharge-pair images for
varying values of $\Delta t$ (indicated on the images), pressure
and voltage pulse amplitude and length (indicated on top). Images
taken in pure nitrogen.}
\end{figure}

A selection of measurement results obtained in pure nitrogen is given
in figure~\ref{fig:Nitrogen-images}. In general, the behaviour of
double-pulse streamers in nitrogen is very similar to that in air
although there are also some notable differences: 
\begin{itemize}
\item As usual, streamers in nitrogen are thinner and branch more than streamers
in air~\cite{Nijdam2010}.
\item The inception cloud in nitrogen is much smaller than in air, therefore
at pressures above 100\,mbar it is not possible to distinguish the
spots on the edge of the inception cloud.
\item The maximum value of $\Delta t$ for which streamer continuation still
occurs (stage (i)) is about 1.5 to 2 times higher than in air. More
details will be given in section~\ref{sec:Duration-of-streamer}. 
\item The value of $\Delta t$ at which new channels start to occur (stages
(iii) and (iv)) is much higher in pure nitrogen than in air. For the
case of 133\,mbar, 8.7\,kV, 400\,ns new channels following the
side of the old ones (stage (iv)) become clearly present around $\Delta t=3$\,\textmu{}s
in air while in pure nitrogen this only starts to occur around $\Delta t=30$\,\textmu{}s. 
\item Even at very high values of $\Delta t$ (10\,ms and above) many second-pulse
streamers still follow the rough paths of first-pulse streamers (stage
(v)). In air this only occurs up to $\Delta t$ = 1\,ms and in general
is less obvious. At $\Delta t$ = 40\,ms in nitrogen the second discharge
seems independent of the first.
\end{itemize}
One thing not shown in figure~\ref{fig:Nitrogen-images}, but very
obvious in the total set of measurement images, is the large variation
from pulse-pair to pulse-pair in pure nitrogen: the value of $\Delta t$
at which a certain phenomenon occurs can vary by about a factor 2
between pulse-pairs while in air the behaviour is much more stable.
This makes a quantitative study of these phenomena in nitrogen more
difficult than in air. Nevertheless it is still possible to see the
trends of streamer morphology as function of time, but quantitative
measurements of properties can easily have error margins of 20\,\%
or more.

\subsection{Pure argon}

\begin{figure}
\includegraphics[width=16cm]{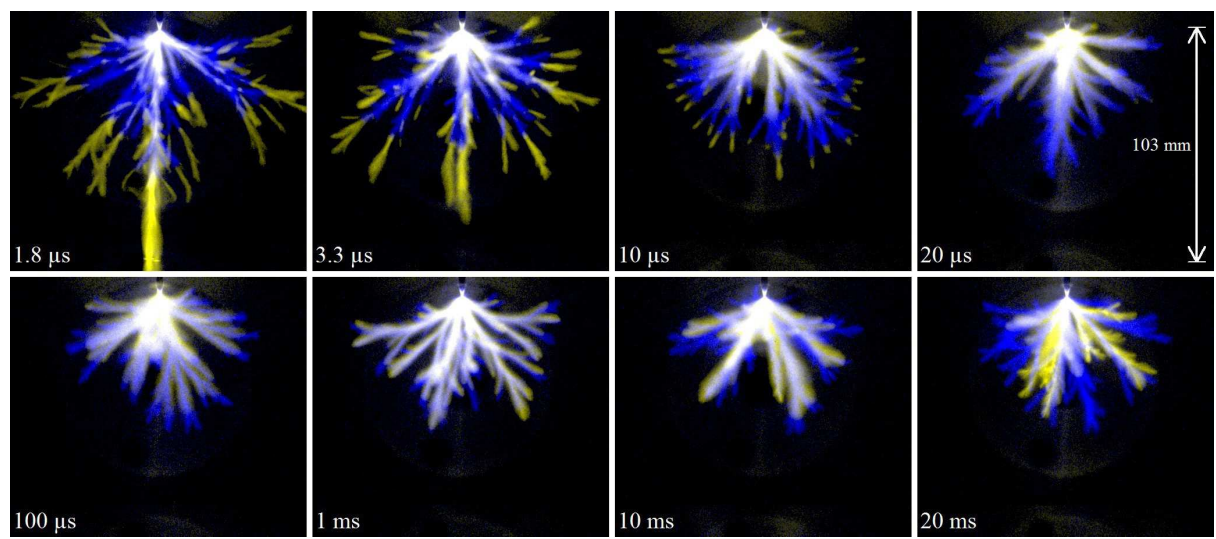}

\caption{\label{fig:OverviewArgon}Superimposed discharge-pair images for varying
values of $\Delta t$ (as indicated in the images). Images taken in
133\,mbar pure argon with pulses of 8.7\,kV amplitude and 200\,ns
length.}
\end{figure}

A third gas we have investigated next to artificial air and pure nitrogen
is argon. The expected purity of argon in our measurements is identical
to that of nitrogen. Because streamer discharges in argon can easily
lead to the occurrence of a spark, the applied voltages and pulse
durations were limited and the lowest pressure used was 133\,mbar.
This prevents the occurence of a spark that can potentially damage
the electrodes as well as the intensifier in the ICCD cameras.

An overview of double-pulse discharges in 133\,mbar argon with a
pulse of 8.7\,kV amplitude and 200\,ns length is given in figure~\ref{fig:OverviewArgon}.
The double-pulse discharges in argon have many similarities with discharges
in air and nitrogen; however there are again some notable differences:
\begin{itemize}
\item The continuation of first-pulse streamers during the second pulse
(stage (i)) takes much longer. While in air and nitrogen the maximum
value of $\Delta t$ for continuation under similar conditions is
less than 2\,\textmu{}s, in argon it is about 15\,\textmu{}s. A
more detailed comparison will be given in section~\ref{sec:Duration-of-streamer}.
\item The decay of excited argon-levels is slower than the decay of the
dominant emitting channels in nitrogen (partly due to energy-pooling
by metastables). Therefore the old channels keep emitting light, even
when no second pulse is given. This means that for lower values of
$\Delta t$ it is impossible to distinguish between light from the
still emitting excited species created during the first pulse and
newly excited species created during the second pulse. This (partly)
explains the large white sections in the centre of the discharge for
$\Delta t\lesssim100$\,\textmu{}s .
\item At values of $\Delta t$ between 100\,\textmu{}s and 1\,ms the second-pulse
streamer channels seem to follow the old channels like in nitrogen
and air (stages (iv) and (v)). However, the large number of channels
makes it difficult to tell whether they exactly follow the old paths
or, like in nitrogen and air, hug the edges of the old paths.
\item The first-pulse discharges show a very feather-like morphology, as
is always visible in argon under similar conditions~\cite{Nijdam2010,Takahashi2011}.
However, up to $\Delta t=10$\,ms the second-pulse streamers (except
for the continuing first-pulse streamers) are very smooth. This fits
well with previous work on the effects of pre-ionization on pure argon
and nitrogen streamer morphology~\cite{Takahashi2011,Nijdam2011c}.
\item Between $\Delta t=100$\,\textmu{}s and $\Delta t=1$\,ms the second-pulse
channels get more narrow, however, between $\Delta t=1$\,ms and
$\Delta t=10$\,ms they become wider again. The first effect can
probably be attributed to a decrease in background-ionization levels
inside the channel, while the second effect may be explained by diffusion
of the remaining (low levels) of background ionization. 
\end{itemize}

\section{Duration of streamer continuation\label{sec:Duration-of-streamer}}

In a more quantitative study of the important time-scales in our experiment,
we have measured the maximum value of $\Delta t$, for which the continuation
of streamers is still observed, or, in other words, the value of $\Delta t$
determining the border between stages (i) and (ii). We label this
quantity $\Delta t_{\mathrm{cm}}$. It was measured as function of
pressure for all gases involved and as function of oxygen concentration
in nitrogen-oxygen mixtures at a fixed pressure. Because the continuation
does not stop instantly at a certain $\Delta t$, $\Delta t_{\mathrm{cm}}$
is defined as the value of $\Delta t$ for which in 50\% of the images
some form of continuation is visible.

Similar parameters for the occurrence of other phenomena as function
of $\Delta t$ have also been considered, most notably the time when
the first new streamers start to occur. However, the jitter of these
phenomena (that happen for larger $\Delta t$) can be as large as
a factor of ten, especially in pure nitrogen. This makes it nearly
impossible to make a coherent data set of such phenomena. 

\begin{figure}
\centering\includegraphics[width=8cm]{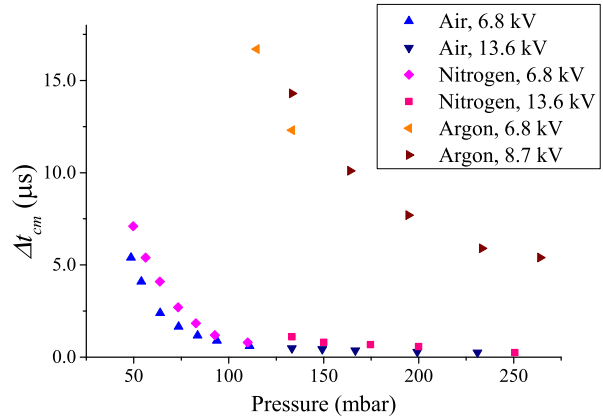}

\caption{\label{fig:Multi-gasses-linear}Maximum pulse-to-pulse interval for
which continuation still occurs ($\Delta t_{\mathrm{cm}}$) in artificial
air, pure nitrogen and pure argon as function of gas pressure for
different pulse voltages as indicated. All measurements were performed
with a pulse length of 300\,ns.}
\end{figure}

Measured values of $\Delta t_{\mathrm{cm}}$ as function of pressure
for three different gases are plotted in figure~\ref{fig:Multi-gasses-linear}.
It was not possible to measure $\Delta t_{\mathrm{cm}}$ over the
entire pressure range with the same voltage pulse amplitude, at lower
pressures a too high voltage pulse would lead to streamers that cross
the gap during the first pulse and therefore no continuation would
be visible. A too low voltage pulse at higher pressures would lead
to no or very short and thin streamers where continuation is difficult
to see. Therefore the applied voltage pulse was varied over the pressure
range, as indicated in the figure.

As could already be concluded from the images shown in the previous
section, $\Delta t_{\mathrm{cm}}$ is much larger for argon than for
nitrogen or artificial air.\\

In order to avoid the effects of changing the reduced electric field ($E/n$)
too much we have also measured $\Delta t_{\mathrm{cm}}$ in artificial
air and pure nitrogen as function of pressure while varying the applied
voltage $V_{max}$ so that $V_{max}/p$ remains constant. We assume
constant temperature so that $n$ scales with $p$ according to the
ideal gas law. Note that this does not imply that everything scales
perfectly with $V_{max}/p$ but just that the total streamer length%
\footnote{Streamer length depends here on streamer velocity and pulse duration.
Because streamer velocity scales with $E/n$, the streamer length
remains roughly constant for a given pulse duration when $V_{max}/p$
is kept constant.%
} remains roughly constant and that streamer inception is ensured.
In this way it is possible to measure $\Delta t_{\mathrm{cm}}$ over
a large pressure range without the need for abrupt changes in applied
voltage.

\begin{figure}
\centering\includegraphics[width=8cm]{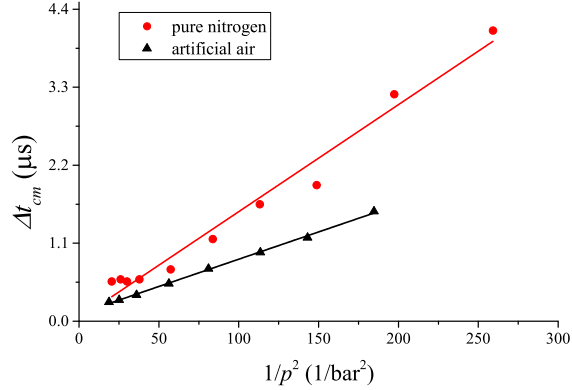}

\caption{\label{fig:dtcm-1overp}$\Delta t_{\mathrm{cm}}$ in artificial air
and nitrogen as function of $1/p^{2}$ with $V_{max}/p$ kept constant
at 77.7 and 74.5\,V/mbar for nitrogen and air respectively. The voltage
pulse length was kept at 200\,ns for both gases. For both gases a
linear fit is included.}
\end{figure}

Unsurprisingly, in the measurements presented in figure~\ref{fig:Multi-gasses-linear}
$\Delta t_{\mathrm{cm}}$ increases with decreasing pressure for all
three gases. The most obvious cause of the stopping of streamer continuation
is a decrease in streamer channel conductivity by various electron
loss processes. Many of the rate-limiting reactions involved in the
electron loss processes are three particle reactions (involving an
electron or an ion and two neutral particles) that scale with $1/n^{2}$,
so we can expect that $\Delta t_{\mathrm{cm}}$ also scales with $1/n^{2}$
with $n$ the neutral density. Therefore we have plotted $\Delta t_{\mathrm{cm}}$
in artificial air and nitrogen as function of $1/p^{2}$ in figure~\ref{fig:dtcm-1overp}.
In section~\ref{sub:Modelling-continuation} we will treat the relevant
electron loss reactions in more detail. 

As was expected we can observe a linear relation between $\Delta t_{\mathrm{cm}}$
and $1/p^{2}$, although the fitted lines do not exactly cross zero.
Unfortunately, because of the tendency of argon discharges to form
a spark, only a limited pressure range could be explored and it was
not possible to measure $\Delta t_{\mathrm{cm}}$ for constant $V_{\mathrm{max}}/p$
.

\subsection{Oxygen concentration effects}

\begin{figure}
\centering\includegraphics[width=8cm]{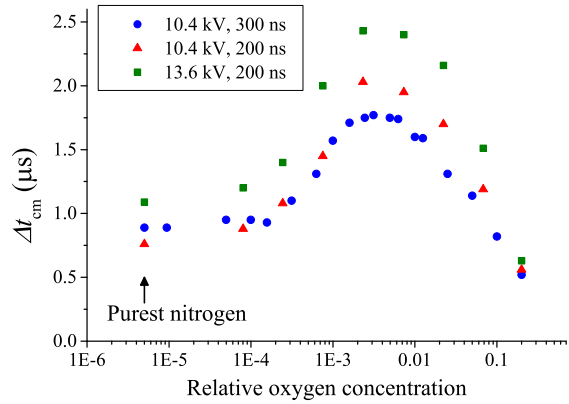}

\caption{\label{fig:dtcm-ratio}$\Delta t_{\mathrm{cm}}$ as function of oxygen
concentration for different oxygen-nitrogen mixtures. All measurements
were performed at 133\,mbar with the pulse amplitudes and durations
as indicated. The points drawn at 5\,ppm oxygen concentration are
actually measured in pure nitrogen gas (within experimental limitations).}
\end{figure}

Figures~\ref{fig:Multi-gasses-linear} and~\ref{fig:dtcm-1overp}
show that although $\Delta t_{\mathrm{cm}}$ is larger for nitrogen
than for artificial air, the difference is relatively small, especially
in comparison with argon. However, if the value of $\Delta t_{\mathrm{cm}}$
would only be influenced by electron attachment to molecular oxygen,
we would expect an increase of $\Delta t_{\mathrm{cm}}$ of at least
a few orders of magnitude between air and pure nitrogen. To investigate
this in more detail, we have measured and plotted $\Delta t_{\mathrm{cm}}$
as function of oxygen concentration. This was done by starting with
artificial air and then repeatedly replacing 50 to 70\,\% of the
gas in the vessel with pure nitrogen, always returning to the starting
pressure to measure $\Delta t_{\mathrm{cm}}$. The results of this
procedure are plotted in figure~\ref{fig:dtcm-ratio}.

We observe that when we decrease the oxygen concentration of artificial
air at 133\,mbar by mixing in more and more pure nitrogen, $\Delta t_{\mathrm{cm}}$
increases as expected. However, this effect reaches a maximum at an
oxygen concentration of about 0.2\,\%. For lower oxygen concentrations
$\Delta t_{\mathrm{cm}}$ decreases again and stabilizes around 0.01\,\%
oxygen. A similar effect has also been found in other streamer measurements:
Ono and Oda~\cite{Ono2003} observed a maximum in time-integrated
discharge current for oxygen concentrations (in nitrogen) around 0.2--1.0\,\%,
not far from the maximum found here although they measured at 1000\,mbar
instead of our 133\,mbar.

Note that although we have calculated from the leak-rate that the
amount of impurities in our purest nitrogen is less than 5\,ppm,
we cannot be entirely sure. It is possible that leaks or impurities
in our gas-handling system lead to higher impurity levels. This would
mean that the leftmost points in figure~\ref{fig:dtcm-ratio} shift
to the right and the plateau below 0.01\,\% oxygen could be an artefact
of our set-up.

\section{Modelling\label{sec:Modelling}}

The development of the discharge during the second pulse depends on
density and distribution of various species (electrons, ions, metastables,
etc.) left over from the first-pulse discharge. If we fix a computational
cell inside of the discharge, neglect drift and diffusion of the species
and follow only the temporal dynamics, we can apply zero-dimensional
modelling to study the left overs of the first-pulse discharge. By
varying the N$_{2}$:O$_{2}$ ratio we can then simulate experimental
results from sections~\ref{sub:Air} and~\ref{sub:Pure-nitrogen}.

\subsection{Model description}

A zero-dimensional model is used to describe the dynamics of species
\begin{equation}
\frac{d[n_{i}]}{dt}=S_{i}\,,\label{eq:kinet}
\end{equation}
 where the source term $S_{i}$ is the total rate of production or
destruction of the species $i$ in various processes; it depends on
the local electric field. A modification of the kinetic file for N$_{2}$-O$_{2}$
mixtures kinetic file from ZDPlasKin~\cite{Zdplaskin,Flitti} is
used that consists of 650 reactions (listed in the supplementary data)
of the 53 species and states listed in table~\ref{tab:Species-considered}.
The plasma chemical processes in N$_{2}$-O$_{2}$ mixtures are mainly
taken from~\cite{Capitelli}. Constant rates for electron-neutral
interactions are calculated using the BOLSIG+ solver that is contained
in the package~\cite{Bolsig}. The list of species and reactions
was automatically converted into the system of ordinary differential
equations (\ref{eq:kinet}) and solved numerically using the ZDPlasKin
tool. The QtPlaskin visualization software was used for the analysis
of the results~\cite{Qtplaskin}.

In order to run ZDPlasKin, the calculation of the electric field has
to be brought into the structure of equation (\ref{eq:kinet}), but
the field profile should also be consistent with the local electron
density. We have solved this problem by 1. neglecting the leading
edge of the ionization front, and by 2. using the conservation of
the total current. 
\begin{equation}
\nabla\cdot(\varepsilon_{0}\ \partial_{t}E+j)=0\,,\label{eq:Max}
\end{equation}
 where $j$ is the electric current; in our case it is $j=-e\ j_{\mathrm{e}}$
with 
\begin{equation}
j_{e}=-\mu(|E|)\ E\ n_{\mathrm{e}}\,,\label{eq:field}
\end{equation}
 where $\mathrm{e}$, $\mu(|E|)$ and $n_{\mathrm{e}}$ are the charge,
electron mobility in N$_{2}$ and density of electrons, respectively.
Electron mobilities are obtained from~\cite{Markosyan,Dujko2011},
where the multi term solution of Boltzmann's equation was used. If
we approximate the weakly curved front by a planar front~\cite{Ebert2011}
and if we assume that the electric field in the non-ionised region
does not change in time ($\partial_{t}E=0$ where $j=0$) , i.e. in
zero-dimensional configuration, we get 
\begin{equation}
\frac{dE}{dt}=\frac{\varepsilon_{0}}{e}\ \mu(|E|)\ E\ n_{\mathrm{e}}\,.\label{eq:formula}
\end{equation}
As said above, we neglect the leading edge of the front and start
integrating at the maximum of the electric field. As~\cite{Naidis2009}
reviews, the maximal electric fields in simulated positive streamers
in STP air vary between 120 and 180\,kV/cm, so we use 150\,kV/cm.
This scales to $E(0)=$20\,kV/cm at 133\,mbar according to the similarity
laws that approximate the fast processes in streamer heads very well.
In the simulations by Naidis~\cite{Naidis2009} we find an electron
density of $5.3\cdot10^{12}$\,cm$^{-3}$ where the electric field
is maximal and scale this to 133\,mbar to use as our initial electron
density $n_{\mathrm{e}}(0)=7.1\cdot10^{11}$\,cm$^{-3}$. We initialise
the densities of the positive ions \foreignlanguage{english}{$\mathrm{N}_{2}^{+}$}
and \foreignlanguage{english}{$\mathrm{O}_{2}^{+}$} such that the
plasma is neutral $n_{\mathrm{e}}(0)=n_{\mathrm{N}_{2}^{+}}(0)+n_{\mathrm{O}_{2}^{+}}(0)$
while keeping the initial ratio $n_{\mathrm{N}_{2}^{+}}(0)$ : $n_{\mathrm{O}_{2}^{+}}(0)$
= $n_{\mathrm{N}_{2}}(0)$:$n_{\mathrm{O}_{2}}(0)$. The initial densities
of the all other ions, excited species and ground-state neutrals (except
N$_{2}$ and O$_{2}$) are assumed to be zero.

This approximation allows us to integrate equations (\ref{eq:kinet})
and (\ref{eq:formula}) together in the form required by ZDPlasKin,
and to get a decay of the electric field that is consistent with the
local conductivity. We integrate the equations until the electric
field reaches 0.67\,kV/cm which is the so-called stability field
for positive streamers in air at 133\,mbar. We remark that though
we have recently criticized this concept in~\cite{Luque:2013ub},
our simulations presented there still show that the average electric
field inside a streamer tree reaches such a value. When the stability
field is reached, we continue to integrate equation (\ref{eq:kinet}),
but now in a constant electric field of 0.67\,kV/cm until the end
of the voltage pulse, and then in a vanishing field.

\begin{table}
\centering

\caption{\label{tab:Species-considered}Species considered in the model.}

\begin{tabular}{l}
\hline 
\tabucline[1.5pt]{-}Ground-state neutrals \tabularnewline
\hline 
N, N$_{2}$, O, O$_{2}$, O$_{3}$\tabularnewline
NO, NO$_{2}$, NO$_{3}$\tabularnewline
N$_{2}$O, N$_{2}$O$_{5}$\tabularnewline
\hline 
\tabucline[1.5pt]{-}Positive ions \tabularnewline
\hline 
N$^{+}$, N$_{2}^{+}$, N$_{3}^{+}$, N$_{4}^{+}$\tabularnewline
O$^{+}$, O$_{2}^{+}$, O$_{4}^{+}$\tabularnewline
NO$^{+}$, N$_{2}$O$^{+}$, NO$_{2}^{+}$, O$_{2}^{+}$N$_{2}$\tabularnewline
\hline 
\tabucline[1.5pt]{-}Excited neutrals \tabularnewline
\hline 
N$_{2}$(A$^{3}$$\Sigma_{\mathrm{u}}^{+}$, B$^{3}\Pi_{\mathrm{g}}$,
C$^{3}\Pi_{\mathrm{u}}$, a$^{\prime}$$^{1}$$\Sigma_{\mathrm{u}}^{-}$)\tabularnewline
N($^{2}$D, $^{2}$P), O($^{1}$D, $^{1}$S)\tabularnewline
O$_{2}$(a$^{1}$$\Delta_{\mathrm{g}}$, b$^{1}$$\Sigma_{\mathrm{g}}^{+}$,
4.5 eV)\tabularnewline
O$_{2}$(X$^{3}$, $v$ = 1 - 4), N$_{2}$(X$^{1}$, $v$ = 1 - 8)\tabularnewline
\hline 
\tabucline[1.5pt]{-}Negative ions \tabularnewline
\hline 
e, O$^{-}$, O$_{2}^{-}$, O$_{3}^{-}$, O$_{4}^{-}$\tabularnewline
NO$^{-}$, NO$_{2}^{-}$, NO$_{3}^{-}$, N$_{2}^{-}$O\tabularnewline
\hline 
\tabucline[1.5pt]{-}\tabularnewline
\end{tabular}
\end{table}

\subsection{Modelling results on streamer continuation\label{sub:Modelling-continuation}}

\begin{figure}
\centering\includegraphics[width=8cm]{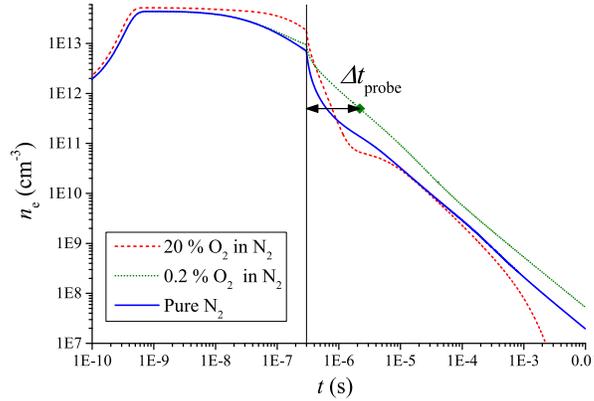}

\caption{\label{fig:ne-development}Modelled development of electron density
as function of time for three different gas mixtures at 133\,mbar.
Pulse duration: 300\,ns, initial electron density: $7.1\cdot10^{11}\,\mathrm{cm^{-3}}$,
maximum electric field on streamer tip: 20\,kV/cm, field in streamer
channel: 0.67\,kV/cm. The end of the pulse is indicated with the
vertical line and the quantity $\Delta t_{\mathrm{probe}}$ is indicated
for the 0.2\,\% oxygen mixture with a probing density of $5\cdot10^{11}\,\mathrm{cm^{-3}}$.}
\end{figure}

\begin{figure}
\centering\includegraphics[width=8cm]{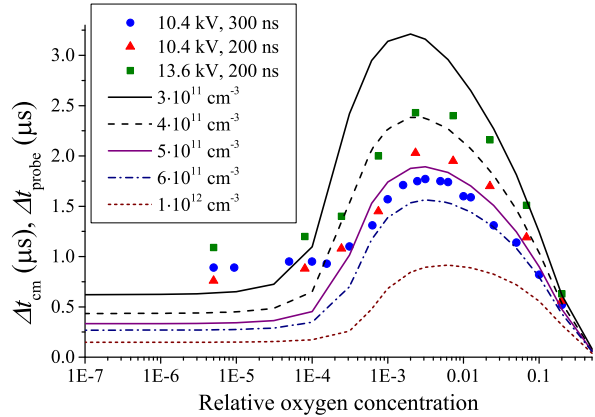}

\caption{\label{fig:dtcm-ratio-model}Measured data from figure~\ref{fig:dtcm-ratio}
combined with modelling results of $\Delta t_{\mathrm{probe}}$ with
varying probe electron densities as indicated in the legend for the
conditions from figure~\ref{fig:ne-development}.}
\end{figure}

The modelled development of $n_{\mathrm{e}}$ as function of time
for artificial air, 0.2\,\% oxygen in nitrogen and pure nitrogen
at 133\,mbar is given in figure~\ref{fig:ne-development}. In this
figure also the quantity $\Delta t_{\mathrm{probe}}$ is indicated
(for the 0.2\,\% oxygen mixture). This quantity represents the time
between the end of the pulse (the moment the field goes to zero) and
the moment a certain probe electron density has been reached. We assume
that the conductivity of the channel depends only on electron density
and that a certain critical conductivity is required for the streamers
to continue their old path during the second pulse. Therefore we can
find this critical density by comparing different $\Delta t_{\mathrm{probe}}$
results with the measured $\Delta t_{\mathrm{cm}}$ values. This is
done in figure~\ref{fig:dtcm-ratio-model}. The shape of the $\Delta t_{\mathrm{cm}}$
and $\Delta t_{\mathrm{probe}}$ curves are remarkably similar: both
rise quickly from artificial air towards decreasing oxygen content,
have a maximum around 0.2\,\% oxygen, decrease for even lower oxygen
concentrations and then level off again around 100\,ppm oxygen. This
indicates that the measured curves are no artefact of the measurement
method, but can be explained by plasma chemical processes. The modelling
results shown for 0.1\,ppm oxygen concentration are virtually identical
to results in pure nitrogen.

Note that except for the chosen electron density probing values, no
fitting parameters were used in the model. Therefore, even this relatively
simple, zero-dimensional model with rate coefficients that can still
have large uncertainties is\textbf{ }able to predict the shape of
$\Delta t_{\mathrm{cm}}$. Only at oxygen concentrations below 100\,ppm
we observe that the model results are always below the measured results.
From the comparison of the model and measurement results we can conclude
that the electron density required for continuation of old streamer
paths is around $5\cdot10^{11}$\,cm$^{-3}$ under the conditions
presented here. 

\begin{figure}
\centering\includegraphics[width=8cm]{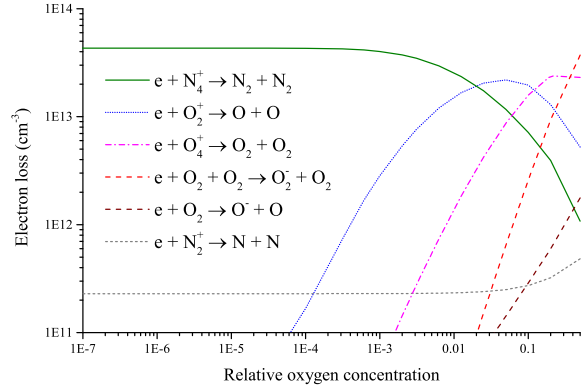}

\caption{\label{fig:Reactionrates}Modelled cumulative electron losses by the
eight dominant electron loss processes as function of oxygen concentration
for the conditions from figure~\ref{fig:ne-development}. Losses
are integrated over time between the start of the simulation and the
moment $n_{\mathrm{e}}$ reaches a probing value of $5\cdot10^{11}\,\mathrm{cm^{-3}}$.}
\end{figure}

The shape of the $\Delta t_{\mathrm{probe}}$ and $\Delta t_{\mathrm{cm}}$
curves can be understood by studying the various electron loss mechanisms.
In figure~\ref{fig:Reactionrates} the cumulative electron losses
by the eight dominant electron loss reactions are plotted as function
of oxygen concentration up to the moment that $n_{e}=5\cdot10{}^{11}\,\mathrm{cm^{-3}}$.
In this figure the dissociative attachment reactions with N$_{2}^{+}$
and O$_{2}^{+}$ leading to ground state atoms are grouped with similar
reactions leading to O($^{1}$D) and N($^{2}$D) states although they
are calculated separately in the actual model.

The figure shows that at low oxygen concentrations (below about 2\,\%)
the electron loss is dominated by dissociative attachment to N$_{4}^{+}$.
At high oxygen concentrations the electron loss is dominated by electron
attachment to molecular oxygen and by dissociative attachment to O$_{4}^{+}$.
All these three processes are relatively fast and thereby can explain
the fast decrease in conductivity at high and low oxygen concentrations.
However, at intermediate oxygen concentrations, a significant part
of the electrons (the major part between 2 and 12\,\% oxygen) are
lost through dissociative recombination with O$_{2}^{+}$, which is
a much slower process than the dominant electron loss reactions at
high and low oxygen concentrations.

\begin{figure}
\centering\includegraphics[width=8cm]{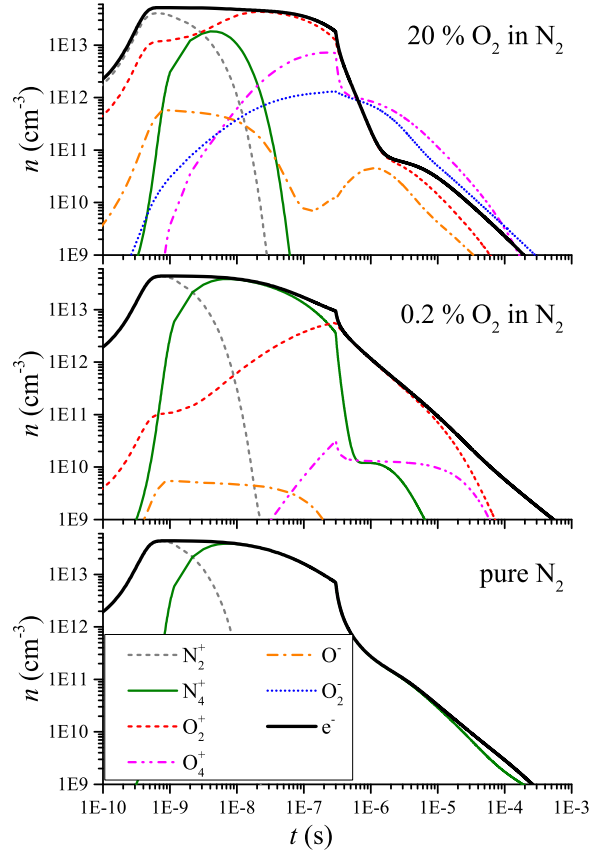}

\caption{\label{fig:Species-development}Modelled development of the densities
of the major ionic species and the electrons in three different gas
mixtures for the conditions indicated in figure~\ref{fig:ne-development}.}
\end{figure}

In air, the main pathway of positive ions is as follows: 
\begin{equation}
\mathrm{N_{2}^{+}\Rightarrow N_{4}^{+}\Rightarrow O_{2}^{+}\Rightarrow O_{4}^{+}},
\end{equation}
as described by Aleksandrov and Bazelyan~\cite{Aleksandrov1999}
and also found in our model results. In this case the electron loss
is dominated by the fast dissociative recombination with O$_{4}^{+}$
as was discussed above. At high nitrogen and oxygen concentrations
this total pathway including recombination is very fast. At low oxygen
concentrations the pathway stops at N$_{4}^{+}$, which also recombines
rapidly with free electrons. However, at intermediate oxygen concentrations
there is enough oxygen to form O$_{2}^{+}$ but not enough to form
reasonable quantities of O$_{4}^{+}$. As, at 133\,mbar, the recombination
rate of electrons with O$_{2}^{+}$ is much lower than with O$_{4}^{+}$,
the total recombination rate is lower at these intermediate concentrations,
which is exactly what we observe in our $\Delta t_{\mathrm{cm}}$
measurements. This is supported by the modelled development of the
major ionic species as is plotted in figure~\ref{fig:Species-development}.

As was shown in section~\ref{sec:Duration-of-streamer}, when $V_{\mathrm{max}}/p$
is kept constant we observe an almost linear relation between $\Delta t_{\mathrm{cm}}$
and $1/p^{2}$. This was attributed to three body processes responsible
for the electron loss. This is not immediately obvious from the list
of dominant reactions in figure~\ref{fig:Reactionrates} where the
attachment of electrons to molecular oxygen is the only three body
reaction. However, also the creation of N$_{4}^{+}$ and O$_{4}^{+}$
requires three body reactions. The dominant production reactions for
these species are: 

\begin{equation}
\mathrm{N_{2}^{+}+N_{2}+N_{2}\rightarrow N_{4}^{+}+N_{2}},
\end{equation}

\begin{equation}
\mathrm{O_{2}^{+}+O_{2}+M_{2}\rightarrow O_{4}^{+}+M_{2},}
\end{equation}
with M either O or N. Therefore it is not surprising that $\Delta t_{\mathrm{cm}}$
scales approximately with $1/p^{2}$. 

\begin{figure}
\centering\includegraphics[width=8cm]{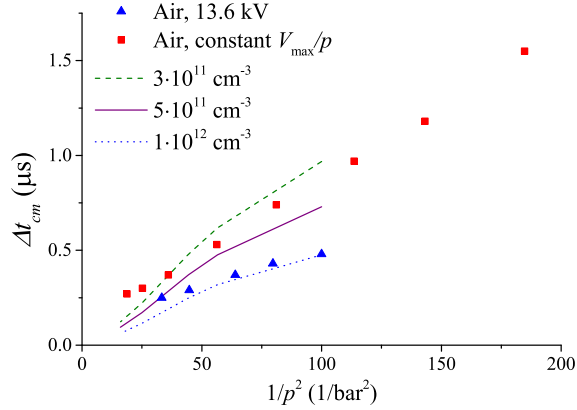}

\caption{\label{fig:Pressure-model}Measured $\Delta t_{\mathrm{cm}}$ as function
of $1/p^{2}$ in artificial air with 13.6 kV, 300\,ns pulses (triangles),
$\Delta t_{\mathrm{cm}}$ at constant $V_{max}/p$ (squares, same
data as plotted in figure~\ref{fig:dtcm-1overp}) compared with modelling
results with 250\,ns pulse lengths for different probe electron densities.}
\end{figure}

However, if we look at the calculated values of $\Delta t_{\mathrm{probe}}$
plotted in figure~\ref{fig:Pressure-model}, there is no perfect
linear relation between $\Delta t_{\mathrm{probe}}$ and $1/p^{2}$.
This may be caused by the constant electric fields that were applied
in the model and that lead to lower than proportional $\Delta t_{\mathrm{probe}}$
for increasing $1/p^{2}$. This can also be observed from the accompanying
measurement results at fixed applied voltage (triangles in the figure).
Unfortunately, the model did not converge for pressures below 100\,mbar.

\subsection{Other modelling results}

The experiments have shown that in 133\,mbar artificial air or pure
nitrogen, new streamers start to occur around $\Delta t\approx2.5$\,\textmu{}s
and $\Delta t\approx35$\,\textmu{}s respectively. From the modelled
development of $n_{\mathrm{e}}$ (see figure~\ref{fig:ne-development})
we cannot fully understand these timescales: there is no probing value
of $n_{\mathrm{e}}$ that gives such a difference in $\Delta t_{\mathrm{probe}}$.
In general the values of $\Delta t_{\mathrm{probe}}$ for artificial
air and pure nitrogen are quite similar for any $n_{\mathrm{e}}$
for probing times below a few hundred microseconds. When looking at
the total ionization density as function of time (i.e. the sum of
electrons and negative ions) the results deviate more from the observations:
the ionization degree is higher in air than in nitrogen at any time
after the pulse. This is because in air the ionization density is
not given by only the electron density but the density of the negative
ions has to be included which increases the ionization density significantly.
Therefore we can conclude that the occurrence of new streamers cannot
be directly linked to either electron or ionization density.

In section~\ref{sub:Pure-nitrogen} we observed that in pure nitrogen
streamers can follow the path of their predecessors even at $\Delta t\geq10$\,ms
while in artificial air this is only observed up to $\Delta t\thickapprox1$\,ms
and even then this occurs much less often. This fits well with the
modelled $n_{\mathrm{e}}$ development which in air starts to drop
much quicker than in nitrogen after about $t=100$\,\textmu{}s. This
also holds for the total ionization degree, which, although (in air)
is higher than the electron density, shows the same trend and drops
below the one for nitrogen at about 1.3\,ms. 

All mentioned numbers are for a pressure of 133\,mbar and room temperature.
Of course electrons will also be lost through (ambipolar) diffusion,
which will start to play a role around millisecond timescales and
is not incorporated in our model.

\section{\label{sec:Discussion-and-conclusions}Summary and conclusions}

\subsection{Continuation of old paths}

In this work we have shown that the application of two consecutive
pulses is a valuable tool to improve our understanding of streamer
discharges and the interaction between repetitive discharges. We have
found that it is possible to \textquotedbl{}reignite\textquotedbl{}
a streamer with a second high voltage pulse so that it continues its
old path. This is only possible for relatively short durations ($\Delta t\leq\Delta t_{\mathrm{cm}}$)
after the first pulse. The measured values of $\Delta t_{\mathrm{cm}}$
are between 0.5 and 15\,\textmu{}s at 133\,mbar depending on the
gas mixture. In nitrogen-oxygen mixtures we have observed that $\Delta t_{\mathrm{cm}}$
exhibits a maximum around 0.2\,\% oxygen in nitrogen. For lower and
higher oxygen concentrations $\Delta t_{\mathrm{cm}}$ decreases.
When modelling the decrease of electron density after a streamer discharge
we find the same pattern. For probe electron densities around $5\cdot10{}^{11}\,\mathrm{cm^{-3}}$
the decay times fit very well with our measurements. Therefore we
conclude that the maximum in $\Delta t_{\mathrm{cm}}$, and the continuation
of old paths in general, can be explained by a loss of conductivity
in the streamer channel due to the loss of free electrons and that,
under our conditions, the minimum $n_{\mathrm{e}}$ for continuation
is about $5\cdot10{}^{11}\,\mathrm{cm^{-3}}$ at 133\,mbar.

The explanation for the maximum in $\Delta t_{\mathrm{cm}}$, or minimum
in total electron loss rate, is that at the oxygen concentrations
of this maximum the major loss process is recombination with O$_{2}^{+}$,
while at lower/higher oxygen concentrations the electron loss is dominated
by recombination with N$_{4}^{+}$ and O$_{4}^{+}$ respectively,
combined with attachment at high oxygen concentrations. These three
processes are all faster than recombination with O$_{2}^{+}$ which
explains the maximum in $\Delta t_{\mathrm{cm}}$.

\begin{figure}
\centering\includegraphics[width=8cm]{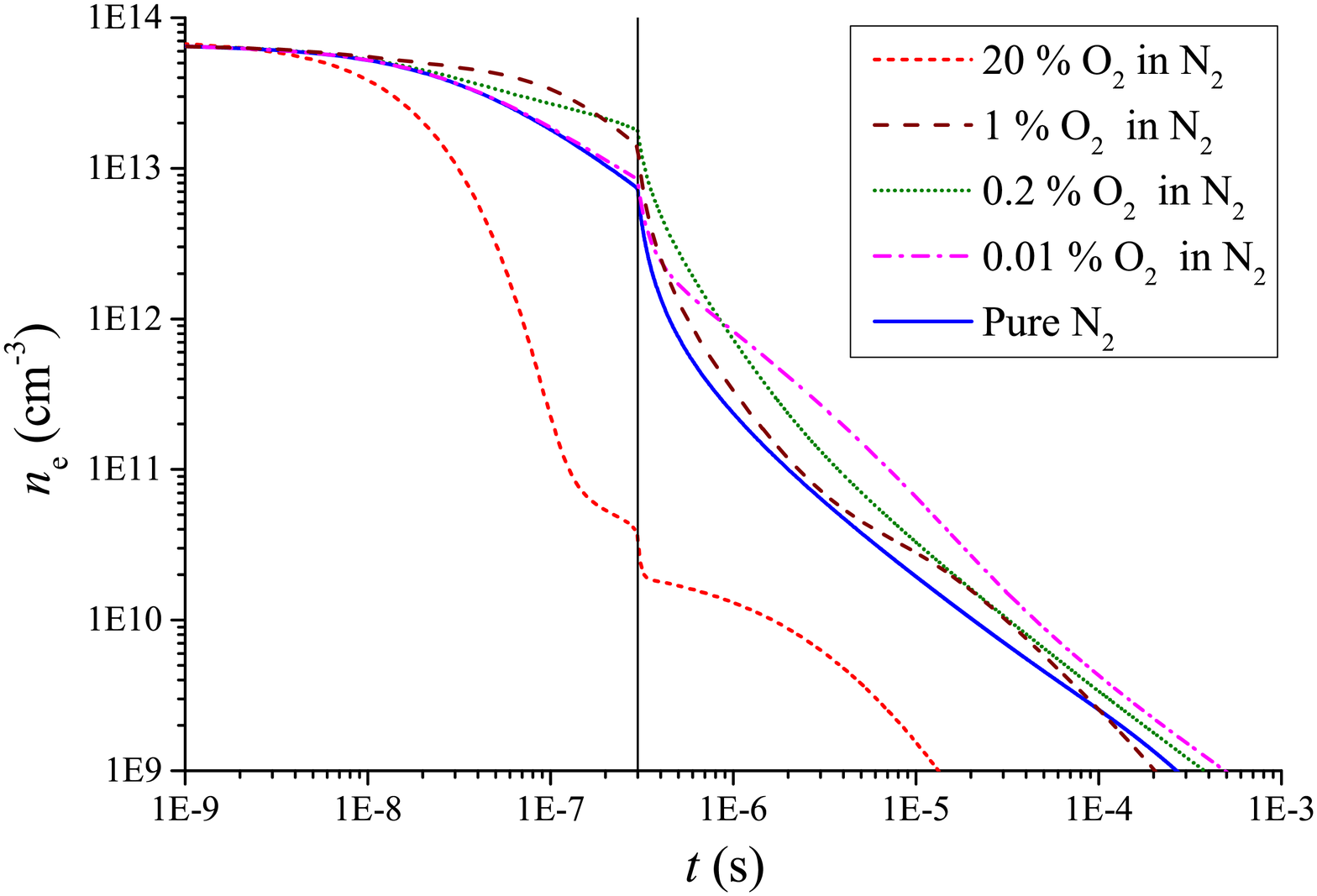}

\caption{\label{fig:ne-atmospheric}Modelled development of electron density
as function of time for five different gas mixtures at 1000\,mbar.
Pulse duration: 300\,ns, initial electron density: $5.3\cdot10^{12}\,\mathrm{cm^{-3}}$,
maximum electric field on streamer tip: 150\,kV/cm, field in streamer
channel: 5\,kV/cm. The end of the pulse is indicated with the vertical
line.}
\end{figure}

Höft \emph{et al~}\cite{Hoft2013} have found an increase in discharge
decay time when changing the oxygen concentration in nitrogen from
20\,\% to 0.1\,\% in a pulsed atmospheric pressure dielectric barrier
discharge. This is in line with our observations. Figure~\ref{fig:ne-atmospheric}
shows that the development of $n_{\mathrm{e}}$ at atmospheric pressure
follows a similar general trend as observed at lower pressures: $n_{\mathrm{e}}$
decays fastest for artificial air, is slower for lower oxygen concentrations,
but is again somewhat faster for pure nitrogen. However, the difference
between air and pure nitrogen is much larger than at lower pressures
due to the much faster three body attachment at atmospheric pressure.
Also the maximum in $\Delta t_{\mathrm{probe}}$ for probe densities
below $1\cdot10^{12}\,\mathrm{cm^{-3}}$ now occurs at 0.01\,\% oxygen,
instead of 0.2\,\% as was found at 133\,mbar%
\footnote{We have calculated the development of $n_{\mathrm{e}}$ at atmospheric
pressure for only five oxygen concentrations so the real maximum can
be higher or lower than 0.01\,\% oxygen.%
}. Höft \emph{et al} conclude that the decay time increases with decreasing
oxygen concentration, and our observations confirm their findings
at atmospheric pressure and oxygen concentrations above 0.01\,\%.
According to our modelling, an extension of these measurements to
higher purity nitrogen should show a maximum in decay time and a decrease
when going to very pure nitrogen.

It is probable that this mechanism can also explain the maximum in
integrated current density (and light emission) found by Ono and Oda~\cite{Ono2003}
at oxygen concentrations around 0.2--1.0\,\% in atmospheric pressure
point-plane discharges. They explain the increase of current when
adding small amounts of oxygen to pure nitrogen by the increased streamer
diameter and photo-ionization. Too much oxygen would lead to electron
attachment and therefore reduces the current. This is in line with
our modelling results, although it disregards the effects of recombination
with O$_{4}^{+}$. The increase in current with small oxygen additions
also matches our modelling results at atmospheric pressure (figure~\ref{fig:ne-atmospheric})
which shows the highest electron density during the pulse ($t<300$\,ns)
for oxygen concentrations of 0.2 and 1\,\%.

Walsh \emph{et al}~\cite{Walsh2012,Naidis2013} also presented a
similar continuation phenomenon in atmospheric pressure plasma-jets
in helium but did not discuss the maximum continuation time as it
was limited by the applied voltage pulse length.

An open question is the exact physical mechanism that determines the
critical or probing density. It is clear that the electron density
at the end of the first voltage pulse is sufficient to let the streamer
return to electrical neutrality between voltage pulses. A possible
hypothesis would be that the probing density is determined by the
Maxwell relaxation time given by the electron density at the beginning
of the second pulse. If that time is comparable to the voltage rise
time, the streamer can be screened again and continue to grow. However,
the Maxwell relaxation times for our conditions are sub-nanosecond
at the electron densities expected at this time (at 133\,mbar), while
the voltage rise time is about 15\,ns. For the same reason one would
expect the critical density to depend on pressure but figure~\ref{fig:Pressure-model}
implies that it is not..

\subsection{Other phenomena}

Besides the continuation of streamers for short pulse-to-pulse intervals,
we have observed a few other interesting phenomena at longer pulse-to-pulse
intervals. When the interval is long enough to prevent streamer continuation
(i.e., the conductivity has dropped enough), the remaining ionization
still prevents new streamers to occur. We attribute this to the too
high electron density which, though too low for continuation, is still
high enough to shield the electrode tip and prevent streamer formation.
During this period we only observe some re-glowing or secondary streamers
in the area of the first discharge, especially at the edge between
inception cloud and streamers as well as at the cathode spots for
the cases where the first streamers have crossed the gap. 

After some time (pulse-pulse intervals of a few to some tens of microseconds,
corresponding to electron densities of $10^{10}-10^{11}$\,cm$^{-3}$
at 133\,mbar) new streamers start to occur. The first of these still
avoid the entire area occupied by the first-pulse discharge. For slightly
larger values of $\Delta t$ the new streamers penetrate the area
of the first-pulse discharge. These new streamers are much thinner
than the first-pulse streamers and follow the edges of the paths of
the original channels. This is similar to the streamers observed in~\cite{Nijdam2011}
that also follow the edge of a pre-ionized region. Apparently the
streamers cannot enter the area of high leftover electron density
of the old channel because they cannot enhance the field there but
preferentially follow the intermediate electron density levels present
at the edge of these channels.

For still larger values of $\Delta t$ (tens to hundreds of microseconds
corresponding to electron densities of $10^{9}-10^{10}$\,cm$^{-3}$
at 133\,mbar) the new streamers grow in size and start to overlap
more with the first-pulse channels. Finally, at even larger values
of $\Delta t$ (electron densities below $10^{8}$\,cm$^{-3}$ at
133\,mbar) they become fully independent of the first-pulse channels.
This transition is very gradual and in most cases some second-pulse
channels still follow a first-pulse channel while others do not. In
nitrogen the following of first-pulse streamers occurs much longer
than in artificial air. This is caused by the faster loss of electrons
in air by electron attachment which, at longer timescales, dominates
over recombination as it scales with $n_{\mathrm{e}}$ instead of
with $n_{\mathrm{e}}n_{+}$.

All these later-stage phenomena can be explained by the same two mechanisms:
streamers prefer and follow elevated levels of electron density but
cannot propagate through areas with too high electron density.

\ack{}{}

SN was supported by FY2012 Researcher Exchange Program between the
Japan Society for the Promotion of Science JSPS and The Netherlands'
Organisation for Scientific Research NWO, ET by JSPS KAKENHI Grant
Number 24560249 and AM by STW-project 10751, part of NWO.

\section*{References}

\bibliographystyle{unsrt}
\bibliography{revised}

\end{document}